\documentstyle[12pt]{article}
\makeatletter

\if@twoside
\else
\fi
\advance\topmargin by -3mm

\ifcase \@ptsize
\or 
\or 
\fi

\def\@citex[#1]#2{%
\if@filesw \immediate \write \@auxout {\string \citation {#2}}\fi
\@tempcntb\m@ne \let\@h@ld\relax \def\@citea{}%
\@cite{%
 \@for \@citeb:=#2\do {%
 \@ifundefined {b@\@citeb}%
 {\@h@ld\@citea\@tempcntb\m@ne{\bf ?}%
  \@warning {Citation `\@citeb ' on page \thepage \space undefined}}%
 {\@tempcnta\@tempcntb \advance\@tempcnta\@ne%
 \@tempcntb\number\csname b@\@citeb \endcsname \relax%
 \ifnum\@tempcnta=\@tempcntb %
 \ifx\@h@ld\relax%
 \edef \@h@ld{\@citea\csname b@\@citeb\endcsname}%
 \else%
 \edef\@h@ld{\ifmmode{-}\else--\fi\csname b@\@citeb\endcsname}%
 \fi%
 \else
 \@h@ld\@citea\csname b@\@citeb \endcsname%
        \let\@h@ld\relax%
      \fi}%
    \def\@citea{,\penalty\@highpenalty\,}%
  }\@h@ld
}{#1}}
\@addtoreset{equation}{section}
\def\section{\@startsection {section}{1}{\z@}{-3.5ex plus -1ex minus
 -.2ex}{2.3ex plus .2ex}{\large\bf\centering}}
\def\subsection{\@startsection{subsection}{2}{\z@}{-3.25ex plus%
 -1ex minus -.2ex}{1.5ex plus .2ex}{\sc}}
\gdef\@publabel{\hfil}
\gdef\@pubdate{\null}
\gdef\@pubnumber{\null}
\gdef\@author{\null}
\gdef\@title{\null}
\gdef\@abstract{\null}
\long\def\pubdate#1{\gdef\@pubdate{#1}}
\long\def\pubnumber#1{\gdef\@pubnumber{#1}}
\long\def\publabel#1{\gdef\@publabel{#1}}
\long\def\author#1{\gdef\@author{#1}}
\long\def\title#1{\gdef\@title{#1}}
\long\def\abstract#1{\gdef\@abstract{#1}}
\def\titlerelax{
\let\maketitle\relax
\let\settitleparameters\relax
\let\consolidatetitle\relax
\let\inittitlepage\relax
\let\finishtitlepage\relax
\let\titlepagecontents\relax
\let\multithanks\relax
\let\titlebaselines\relax
\let\@makepub\relax
\let\@maketitle\relax
\let\@makeauthor\relax
\let\@makeabstract\relax
\let\@maketitlenote\relax
\let\thanks\relax
\let\titlerelax\relax}
\def\titleclean
{\gdef\@titlenote{}
\gdef\@abstract{}
\gdef\@author{}
\gdef\@title{}
\gdef\@pubdate{}\gdef\@pubnumber{}\gdef\@publabel{}
\gdef\@dpublabel{}
}

\def\@makepub{\vbox to \z@{\hbox to \textwidth{\hfill
\@publabel \hfill
\llap{\parbox[t]{0.33\textwidth}{\raggedleft\@pubnumber}}}%
\vss}}
\def\@maketitle{\vskip 60pt \begin{center}
 {\LARGE \@title \par}
 \end{center}}

\def\@makeauthor{{%
\def\and{\smallskip {\normalsize \rm and\smallskip }}
\def\And{\medskip {\normalsize \rm and\\}\medskip}
\long\def\address##1{{\def\and{\\and\\}\medskip
				{\small \it \\##1\\}
}}
{\centering
 \vskip 3em
 \large \lineskip .75em
 \@author}
 \par}}
\def\@makedate{\vskip 1.5em
 {\raggedright \small \noindent\@pubdate \par}}
\def\@makeabstract{\vskip 1.5em
{\small
\begin{center}
{\bf ABSTRACT\vspace{-.5em}\vspace{0pt}}
\end{center}
\quotation \@abstract \endquotation}}
\def\maketitle{\titlepage
\let\footnotesize\small \setcounter{page}{0}
\@makepub
\vfil
\@maketitle
\@makeauthor
\vfil
\@makeabstract
\@thanks
\vfil
\@makedate
\if@restonecol\twocolumn \else \eject \fi
\titlerelax \titleclean
\setcounter{footnote}{0}
}

\def\bigans{y }

\read-1 to\answ
\ifx\answ\bigans
\read-1 to\bnsw
	\ifx\bnsw\bigans 

\ifcase\@ptsize
 \font\tenmsa=msam10
 \font\sevenmsa=msam7
 \font\fivemsa=msam5
 \font\tenmsb=msbm10
 \font\sevenmsb=msbm7
 \font\fivemsb=msbm5
\or
 \font\tenmsa=msam10 scaled \magstephalf
 \font\sevenmsa=msam8
 \font\fivemsa=msam6
 \font\tenmsb=msbm10 scaled \magstephalf
 \font\sevenmsb=msbm8
 \font\fivemsb=msbm6
\or
 \font\tenmsa=msam10 scaled \magstep1
 \font\sevenmsa=msam8
 \font\fivemsa=msam6
 \font\tenmsb=msbm10 scaled \magstep1
 \font\sevenmsb=msbm8
 \font\fivemsb=msbm6
\fi

\newfam\msafam
\newfam\msbfam
\textfont\msafam=\tenmsa  \scriptfont\msafam=\sevenmsa
  \scriptscriptfont\msafam=\fivemsa
\textfont\msbfam=\tenmsb  \scriptfont\msbfam=\sevenmsb
  \scriptscriptfont\msbfam=\fivemsb

\def\hexnumber@#1{\ifnum#1<10 \number#1\else
 \ifnum#1=10 A\else\ifnum#1=11 B\else\ifnum#1=12 C\else
 \ifnum#1=13 D\else\ifnum#1=14 E\else\ifnum#1=15 F\fi\fi\fi\fi\fi\fi\fi}

\def\msa@{\hexnumber@\msafam}
\def\msb@{\hexnumber@\msbfam}

\def\Bbb{\ifmmode\let\next\Bbb@\else
 \def\next{\errmessage{Use \string\Bbb\space only in math mode}}\fi\next}
\def\Bbb@#1{{\Bbb@@{#1}}}
\def\Bbb@@#1{\fam\msbfam#1}
	\else
		\ifx\cnsw\bigans 

\ifcase\@ptsize
 \font\tenmsx=msxm10
 \font\sevenmsx=msxm7
 \font\fivemsx=msxm5
 \font\tenmsy=msym10
 \font\sevenmsy=msym7
 \font\fivemsy=msym5
\or
 \font\tenmsx=msxm10 scaled \magstephalf
 \font\sevenmsx=msxm8
 \font\fivemsx=msxm6
 \font\tenmsy=msym10 scaled \magstephalf
 \font\sevenmsy=msym8
 \font\fivemsy=msym6
\or
 \font\tenmsx=msxm10 scaled \magstep1
 \font\sevenmsx=msxm8
 \font\fivemsx=msxm6
 \font\tenmsy=msym10 scaled \magstep1
 \font\sevenmsy=msym8
 \font\fivemsy=msym6
\fi

\newfam\msxfam
\newfam\msyfam
\textfont\msxfam=\tenmsx  \scriptfont\msxfam=\sevenmsx
  \scriptscriptfont\msxfam=\fivemsx
\textfont\msyfam=\tenmsy  \scriptfont\msyfam=\sevenmsy
  \scriptscriptfont\msyfam=\fivemsy

\def\hexnumber@#1{\ifnum#1<10 \number#1\else
 \ifnum#1=10 A\else\ifnum#1=11 B\else\ifnum#1=12 C\else
 \ifnum#1=13 D\else\ifnum#1=14 E\else\ifnum#1=15 F\fi\fi\fi\fi\fi\fi\fi}

\def\msx@{\hexnumber@\msxfam}
\def\msy@{\hexnumber@\msyfam}

\def\Bbb{\ifmmode\let\next\Bbb@\else
 \def\next{\errmessage{Use \string\Bbb\space only in math mode}}\fi\next}
\def\Bbb@#1{{\Bbb@@{#1}}}
\def\Bbb@@#1{\fam\msyfam#1}

\else
\def\Bbb#1{{\bf #1}}
		\fi
	\fi
\else
\def\Bbb#1{{\bf #1}}
\fi

\def\cF{{\cal F}}
\def\cG{{\cal G}}
\def\cJ{{\cal J}}

\def\cN{{\cal N}}

\let\s=\sigma
\def\blank#1{}
\def\cev#1{\langle #1 \vert}
\def\cont{\nonumber\\*&&\mbox{}}
\def\D#1{ \frac{d#1}{2\pi i}}
\def\en{\end{equation}}
\def\eq{\begin{equation}}
\def\eqq{\begin{eqnarray}}
\def\enn{\end{eqnarray}}
\def\id{\bf 1}
\def\tL{l}
\def\tG{g}
\def\vac{\vec 0}
\def\vec#1{\vert #1 \rangle}

\relax

\begin{document}

\def\vf{\vfill\eject}
\def\vf{}
\pubnumber{DAMTP 93--14}
\pubdate{6 June 1993}
\title{Null Vectors of the Superconformal Algebra:\\
 The Ramond Sector}
\author{G.~M.~T.~WATTS$^{1}$
\address{~\\
{}~\\
St.\ John's College,\\
St.\ John's Street,\\
Cambridge, CB2 1TP, U.K.\\
{}~ \\
and\\
{}~ \\
Department of Applied Mathematics and Theoretical Physics,\\
University of Cambridge,
Silver Street,\\
Cambridge, CB3 9EW, U.K.
}
}

\footnotetext[1]{
Email: {\tt G.M.T.Watts@amtp.cam.ac.uk}
}

\abstract{%
We  consider the Ramond sector of the $N=1$ superconformal
algebra and find expressions for the singular vectors in reducible
highest weight Verma module representations by the fusion principle of
Bauer et al.}

\maketitle

\def\all#1#2{#2}

\section{Introduction}
\label{sec.intro}

Conformal field theories in two dimensions describe statistical
systems at a second order phase transition. The study of these field
theories may be manageable because the
infinite dimensional symmetry algebras which exist can reduce the field
content to a finite number of representations. In such minimal cases,
there are `null vectors' which should decouple from all correlation
functions. As a result the non-zero  correlation functions satisfy
differential equations  which enable one to solve the theory
completely. It is  an
interesting problem to study the null vector structure in conformal
models.
The simplest infinite dimensional algebra which arises is the algebra
of conformal transformations itself, the Virasoro algebra. Explicit
formulae for a restricted class of Virasoro null vectors were found by
Benoit and Saint-Aubin \cite{BStA1} and extended to all cases by Bauer
et al. \cite{BDIZ1,BDIZ2} and by Kent \cite{Kent5}.
The next simplest algebra is the algebra of superconformal
transformations, and again null vectors play an important r\^ole in
determining the structure of the theory \cite{FQSh3}. There are two
sectors in superconformal field theory, determined by the boundary
conditions on the fermionic fields, the Neveu-Schwarz and the Ramond,
or the NS and R respectively.
A subclass of the NS null vectors were again found by Benoit and
Saint-Aubin \cite{BStA3,BStA4} and then extended to all NS cases using
methods similar to those  of Bauer et al. by Benoit and Saint-Aubin
\cite{BStA2} and by
Hwang et al. \cite{HZZh1,HZZh2}. In this paper we shall use use similar
methods again to give formulae for \all{all }{a class of }  Ramond
Algebra null vectors\all{.}{ and conjecture formulae for the rest.}

The paper is arranged as follows.
In section \ref{sec.rep}
we review the NS and R representation theory.
In section \ref{sec.field}
we describe our conventions for the transformation properties of the
fields under superconformal transformations.
In section \ref{sec.ope} we consider the definition of the operator
product expansion, and show that the presence of a null state in one
of the two representations being fused leads to recursive equations
for the states appearing in the OPE (the descent equations)
and the possibility of finding new null states.
We apply this to the same simple null vectors as before
in section \ref{sec.ope.null}
and using this we obtain
in section \ref{sec.p1}
an explicit formula for the $(2n,1)$  class of Ramond null
vectors.
In section \ref{sec.pq}
we present \all{and prove}{} a formula for all Ramond null
vectors.
Section \ref{sec.conc}
contains come concluding remarks.
\section{Representation theory}
\label{sec.rep}

The algebra of chiral superconformal transformations in the plane is
generated by
two fields, $L(z) = \sum_m L_m z^{-m-2}$ and $G(z) = \sum_m G_m
z^{-m-3/2}$. According to the choice of boundary conditions on $G(z)$,
one may choose the labels $m$ on $G_m$ to be integer or half-integer;
the algebra in these two sectors are called the Ramond and Neveu
Schwarz algebras respectively.
In both sectors the generators obey
\eqq
{}~[L_m,L_n]  &=&
\frac c{12}	 m(m^2-1)\delta_{m+n} + (m-n)L_{m+n}  \;,\\
 \{G_m,G_n\} &=& \frac c3
		(m^2 - \frac 14) \delta_{m+n} + 2 L_{m+n} \;,\\
{}~[L_m,G_n]  &=& (\frac m2 - n)G_{m+n}\;.
\enn
In a superconformal field theory the physically relevant
representations of the superconformal algebra are irreducible highest
weight representations. These are graded by $L_0$ eigenvalue, or level.

\subsection{NS representations}

Highest weight representations (hwrs) of the NS algebra have a state
of least $L_0$ eigenvalue $\vec h$ such that
\eqq
L_m |h\rangle &=& h \delta_m |h\rangle,\; m\ge 0
\label{eq.hw.l} \;, \\
G_m |h\rangle &=& 0,\; m\ge 1/2
\label{eq.hw.g} \;.
\enn
A minimal set of conditions is that
$G_{1/2} \vec h = G_{3/2} \vec h = 0$; the others follow from the 
algebra relations.
The Verma module $M_h$ has a basis
\eq
L_{i_1}\ldots L_{i_m} G_{j_1}\ldots G_{j_n}\vec h \l,
\label{eq.verma}
\en
where
$i_p \le i_{p+1}<0$,  and $ j_p < j_{p+1}<0$. We denote the subspace of
$M_h$ of $L_0$ eigenvalue $(h+n)$ by $M_h^{(n)}$. We say that states
in $M_h^{(n)}$ are at level $n$ in $M_h$.
The irreducible highest weight representation $L_h$ is the quotient of
$M_h$ by its maximal invariant submodule.
Again we denote the subspace of $L_h$ of $L_0$
eigenvalue $(h+n)$ by $L_h^{(n)}$.
The Verma module $M_h$ contains a highest weight state at level
$pq/2$
whenever $h=h_{p,q}(t)$ and  $c=c(t)$
where
\eq
c(t) = 15/2 - 3/t - 3t \;,
\label{eq.c.t}
\en
and
\eq
h_{p,q}(t) = (1-pq)/4 + (q^2 - 1)t/8 + (p^2-1)/(8t)  \;, 
\en
and $p,q\in\Bbb Z$, $p+q \in 2\Bbb Z$ (see refs.\ \cite{Kac2,FeFu1,FeFu2}).
Explicit formulae for the null vectors of type $(2p+1,1)$, $(1,2p+1)$
were given in \cite{BStA3} and of type $(p,q)$, $p-q$ even in
\cite{BStA2,HZZh1}.

\subsection{R representations}

Highest weight representations of the Ramond algebra
have a state $\vec\lambda$ of least $L_0$ eigenvalue
satisfying
\eq
\begin{array}{rcll}
L_m |\lambda\rangle &=& (\lambda^2 + c/24) \delta_m \vec\lambda
\;, & m\ge 0 \nonumber \;,\\
G_m |\lambda\rangle &=& \lambda \delta_m|\lambda\rangle
\;, & m\ge 0 \;,
\end{array}
\label{eq.hw.g.r}
\en
A minimal set of conditions for $\vec\lambda$ to be a Ramond highest
weight state is $( G_0 - \lambda)\vec\lambda = L_1\lambda = 0$; the
others follow from the algebra relations.
The Verma module $M_\lambda$ is again spanned by state of the form
(\ref{eq.verma}).
These too have highest weight states at levels  $pq/2$,
whenever $\lambda=\lambda_{p,q}(t)$ and  $c=c(t)$
where $c(t)$ is given by (\ref{eq.c.t}),
\eq
\lambda_{p,q}(t) =
{{ p - qt }\over{ 2 \sqrt{ 2 t }}} \;,
\en
and $p,q\in \Bbb Z$, $p-q$ an odd integer (see refs.\ \cite{MRCa1,FQSh3}).
The null state at level $pq/2$ has
$G_0$ eigenvalue (see ref. \cite{CFri1})
\eq
\lambda'=
(-1)^q {{ p + qt }\over{ 2 \sqrt{ 2 t }}} \;.
\label{eq.r.gonull}
\en

In this paper we give explicit formulae for
Ramond null vectors $(2p,1)$, $(1,2p)$%
\all{, and a recursive algorithm to calculate general Ramond null
vectors, which could also be written in an explicit, if cumbersome,
form if so desired}{.}

\vf

\section{Field representations}
\label{sec.field}

To define correlation functions which are covariant under
(super)conformal transformations, it is necessary to consider how
fields transform.
 We shall denote
the field corresponding to the state $\vec \psi$ by $\psi(z)$, and
define $\psi(z) \vac = \exp(z L_{-1} ) \vec \psi$.
We first review the commutation relations with $L_m$, and then
consider the relations with $G_m$ for NS fields and R fields in turn.
The relations in this section are only meant to be heuristic -- they
are only used to inspire the definition of an operator product
expansion in section \ref{sec.ope}.

\subsection{Virasoro transformation properties}

We have
\eq
L(z)\vec\psi = \frac{h}{z^2}\vec\psi + \frac 1z L_{-1}\vec\psi + O(1)
\;,
\label{eq.l.c1}
\en
and so we obtain the operator product expansion.
\eq
L(z) \psi(\zeta) = \frac{ h \psi(\zeta) }{(z-\zeta)^2}
+ \frac{\partial\psi(\zeta)}{(z-\zeta)} + O(1) \qquad |z|>|\zeta|
\;.
\label{eq.l.c2}
\en
We can also define the continuation of the ope, in the sense of
analytic continuation inside correlation functions, as
\eq
L(z) \psi(\zeta) \qquad  {}_{\Big(\, |z|>|\zeta| \, \Big) }\;
=
\psi(\zeta) L(z) \qquad  {}_{\Big(\, |\zeta|>|z|\, \Big) }\;.
\label{eq.l.c3}
\en
Then, by taking a contour integral we find
\eqq
L_m \psi(\zeta) &=&
	\oint_{0, |z|>|\zeta|} z^{m+1} L(z)\psi(\zeta) \D z \nonumber\\
&=& \oint_{\zeta} z^{m+1} L(z)\psi(\zeta) \D z +
	\oint_{0, |\zeta|>|z|} z^{m+1} L(z)\psi(\zeta) \D z \nonumber\\
&=& \zeta^{m}( h(m+1) + z\partial )\psi(\zeta) + \psi(\zeta)L_m
\;,
\label{eq.l.c4}
\enn
as usual. Following Feigin and Fuchs, and for later convenience let us
define
\eq
\tL_m(z) = L_m - z L_{m-1}
\;,
\label{eq.tl.def}
\en
then
\eq
 ~[\tL_m, \psi(z) ] = h z^m \psi(z)
\;.
\en
\blank{
\eqq
\tL_m(z) &=& L_m - z L_{m-1} \nonumber\\
e_m(z) &=& \tL_m - z \tL_{m-1}  \qquad  m\le -2 \nonumber\\
e_0'(z) &=& \tL_0 - z \tL_{-1}
\label{eq.tl.def}\\
e_0''(z) &=& -z \tL_{-1}  \nonumber
\enn
then
\eq
 ~[\tL_m, \psi(z) ] = h z^m \psi(z) \quad,\qquad
 ~[ e_m, \psi(z)] = [e_0', \psi(z)] = 0 \quad,\qquad
 ~[ e_0'', \psi(z)] = -h \psi(z)
\en
}
Note that the operators $\tL_m$ generate a closed Lie algebra with
central extension.

\subsection{NS fields and $G_m$}

The action of $G(z)$ on a NS highest weight state $\vec\psi$ is
\eq
G(z) \vec\psi = \frac 1z G_{-1/2}\vec\psi + O(1)
\;,
\label{eq.ns.c1}
\en
and so the operator product expansion is
\eq
G(z) \psi(\zeta) = \frac 1{z-\zeta} \hat G_{-1/2}\psi(\zeta) + O(1)
\qquad  \Big(\,|z|>|\zeta|  \, \Big)
\;,
\label{eq.ns.c2}
\en
where the field $\hat X_{-n}\phi(z)$ is defined for any field $\phi$
and mode $X_n$ via
\eq
lim_{z \to 0}  \hat X_{-n}\phi(z) \vac = X_{-n}\vec\phi
\;.
\en
Although equation (\ref{eq.ns.c1}) is only defined for $G$ acting in
the NS sector, we shall assume (\ref{eq.ns.c2}) is valid for $G$ in
both the NS and the R sectors.

Similarly we have
\eq
G(z) \hat G_{-1/2}\psi(\zeta) =
\frac{ 2h}{(z-\zeta)^2} \psi(\zeta)
+ \frac 1{z-\zeta} \partial\psi(\zeta) + O(1)
\qquad
|z|>|\zeta|
\;.
\label{eq.ns.c3}
\en
To continue $G(z) \psi(\zeta)$ to values $|\zeta|>|z|$, we have to
allow for $\psi(\zeta)$ both bosonic and fermionic (consider e.g. the
$c=3/2$ model where both cases arise) and so
\eq
G(z) \psi(\zeta) \qquad  {}_{\Big(\, |z|>|\zeta|  \, \Big) }\;
=
\eta_{\psi}
\psi(\zeta) G(z) \qquad  {}_{\Big(\, |\zeta|>|z|  \, \Big) }
\;,
\label{eq.ns.c4}
\en
where $\eta_\psi $ is $1$ for $\psi$ bosonic and $-1$ for fermionic.
Then as with eqn.\ \ref{eq.l.c4} we have
\eqq
G_m \psi(\zeta) &=& \zeta^{m+1/2} \hat G_{-1/2}\psi(\zeta)
+ \eta_\psi \psi(\zeta) G_m
\nonumber\\
G_m \hat G_{-1/2}\psi(\zeta) &=&
	\zeta^{m-1/2}( h(2m+1) + \zeta\partial) \psi(\zeta)
	- \eta_\psi \hat G_{-1/2}\psi(\zeta)
\;,
\label{eq.ns.c5}
\enn
and so if we define
\eq
\tG_m(z) =  G_m - z G_{m-1}  
\;,
\label{eq.tg.def}
\label{eq.ns.c6}
\en
 we have
\eq
\tG_m(z) \psi(z) = \eta_\psi \psi(z) \tG_m(z)
\;.
\label{eq.ns.c7}
\en

\subsection{R fields and $K_m$}
\label{ssec3.r}

Equation \ref{eq.ns.c5} expresses the commutation of a mode $G_m$ with
a Neveu-Schwarz field. For Ramond fields, we find that there is a cut
in the operator expansion of $G(z)$ and $\phi_\lambda(\zeta)$, and as
a result a Ramond field changes the moding of the field $G(z)$ from
integral to half-integral and vice-versa. There is certainly no
possibility of obtaining a commutation relation between $G_m$ and
$\phi_\lambda(\zeta)$, as $G_m$ for fixed $m$ is only defined on one
side,  on the left or the right of $\phi$. As a consequence we shall
instead
	%
follow Friedan et al. ref.\ \cite{FQSh3}, and introduce infinite
linear sums of modes $G_m$, which we write
$K_m,K^+_m$ and find their exchange relations with the Ramond primary
fields. (Friedan et al.\  used
an extended superconformal algebra containing the element $(-1)^F$
where $F$ is the fermion number operator, and a different field basis
as a result, ans our results are not directly comparable with theirs.)

We have
\eqq
G(z) \vec\lambda &=&
\left(\quad
\frac{\lambda}{z^{3/2}} + \frac{G_{-1}}{z^{1/2}}
\;,
\quad\right)\quad
\vec\lambda
\label{eq.r.c1}\\
G(z) \psi_\lambda(\zeta) &=&
\frac{\lambda\psi_\lambda(\zeta)}{(z-\zeta)^{3/2}}
+ \frac{\hat G_{-1}\psi_\lambda(\zeta)}{(z-\zeta)^{1/2}} +
O((z-\zeta)^{1/2})
\qquad  \Big(\,  |z|>|\zeta| \, \Big)
\;.
\label{eq.r.c2}
\enn
To continue eqn.\ \ref{eq.r.c2} to values $|\zeta|>|z|$, consider
\eq
G(z) \psi_\lambda(\zeta) f(z - \zeta)
\qquad {}_{\Big(\, |z|>|\zeta| \, \Big)} \;
=
\psi_\lambda(\zeta) G(z) f(\zeta - z)
\qquad   {}_{\Big(\, |\zeta|>|z| \, \Big) }
\;.
\label{eq.r.c3}
\en
We certainly need $f(z)$ to have a branch cut at the origin if
eqn.\ \ref{eq.r.c3} is to be single valued, but the choice is linked
to the choice of adjoint on fields.
If $(\psi_\lambda)^\dagger = \psi_{\lambda} $ then $f(z) = z^{3/2}$ is a
good choice. If $(\psi_\lambda)^\dagger = \psi_{-\lambda} $
then $f(z) = z^{1/2}$ is a good choice.
So, if we put $(\psi_\lambda)^\dagger = \psi_{\epsilon\lambda}$ we
have
\eq
G(z) \psi_\lambda(\zeta) (z - \zeta)^{3/2}
\qquad   {}_{\Big(\, |z|>|\zeta| \, \Big) } \;
=
\epsilon \psi_\lambda(\zeta) G(z) (\zeta - z)^{3/2}
\qquad   {}_{\Big(\, |\zeta|>|z| \, \Big) }
\;.
\label{eq.r.c4}
\en
However, since we shall use eqn.\ \ref{eq.r.c4} to define the fusion,
and hence the choice of adjoint, this is really a consistency check.

Following Friedan et al.\ ref.\ \cite{FQSh3}, define
\eqq
K_m^+(\zeta) &=&
	\oint_{0, |z|>|\zeta|} z^m (z-\zeta)^{1/2}G(z) \D z
\nonumber\\
    &=& G_m -\frac{\zeta}{2} G_{m-1} -\frac{\zeta^2}8 G_{m-2}+\ldots
\label{eq.r.c5a}	\\
K_m(\zeta) &=&
	\zeta^{-1/2} \oint_{0, |\zeta|>|z|} z^m(\zeta-z)^{1/2}G(z)\D z
\nonumber\\
    &=& G_m - \frac{1}{2\zeta}G_{m+1} - \frac{1}{8\zeta^2}G_{m+2}+\ldots
\;,
\label{eq.r.c5b}
\label{eq.r.c5}
\enn
then
\eq
(K_m( z ))^\dagger = K_m^+(\frac{1}{z^*})
\;,
\label{eq.r.c6}
\en
and we obtain the desire relation
\eq
K_m^+(\zeta) \psi_\lambda(\zeta)
=
\lambda \zeta^m \psi_\lambda(\zeta)
- \epsilon_\lambda \psi_\lambda(\zeta) K_{m-1/2}(\zeta) \zeta^{1/2}
\;,
\label{eq.r.c7}
\en
where $m$ can be integral or half-integral as circumstances dictate.

\vf

\section{Operator product expansions}
\label{sec.ope}


Physicists are used to thinking of defining fusion in terms of an
operator product expansion, that is a map from two representations
into a third (see e.g. \cite{BPZ}), rather than as a map from three
representation spaces to  $\Bbb R$ or $\Bbb C$ (see e.g.
\cite{FeFu3}).
We shall restrict ourselves to  operator product expansions of the
form $\phi_2(z) \vec\psi$ where $\phi_2(z)$ is a field corresponding
to a highest weight state of the superconformal algebra, and
$\vec\psi \in M_3$, $\phi_2(z) \vec\psi\in M_1$ where $M_1$ and $M_3$
are Verma  modules of the superconformal algebra. Rather than attempt
a definition of the operator $\phi_2(z)$ in this  situation,
we consider an OPE to be a collection of maps of Verma modules, $M_3
\to M_1$, $\vec\psi
\to f_n(\psi)$
such that
\eq
\phi_2(z) \vec\psi = \cF(z) \equiv \sum_{n \ge 0} z^{h_1-h_2-h_3+n}
f_n(\psi)
\;,
\label{eq.ope1}
\en
obeys the covariance properties under the superconformal algebra that
one would expect from
eqns.\ \ref{eq.hw.l},\ref{eq.hw.g},\ref{eq.hw.g.r}, \ref{eq.l.c4} and
\ref{eq.ns.c5}.

It was the observation of Bauer et al.\ that for the Virasoro algebra
one could use knowledge of null states in $M_3$ either to determine
$f_n(\vec 3)$ for all $n$ for the highest weight state $\vec 3\in L_3$
or to determine $f_n(\vec 3)$ for $0 \leq n < N$ and obtain an
explicit expression for a null state in $M_1$ at level $N$. Their
method resulted in expressions for the general Virasoro null vector
(refs.\ \cite{BDIZ1,BDIZ2}) and has been subsequently applied to
generate expressions for all NS highest weight null vectors
(refs.\ \cite{BStA3,HZZh1}) and a class of highest weight null vectors
of the $W_3$ algebra (ref.\ \cite{BWat2}).
It will be instructive to repeat the arguments of Bauer et al.\ here
briefly.

\subsection{Virasoro case}
\label{ssec6.vir}

The Virasoro algebra representation theory is simpler than that of the
superconformal algebra. Highest weight representations have a state
$\vec h$ which obeys eqn.\ \ref{eq.hw.l},
and highest weight Verma modules $M_h$ are spanned by states of the form
(\ref{eq.verma}) omitting the modes of $G$. There is highest weight
state in $M_h$ at level $pq$ whenever
$c = c(t) = 13 - 6t - 6/t$ and $h=h_{p,q}= ((p-qt)^2 - (1-t)^2)/4t$
for $p,q\in\Bbb Z$
(see e.g. refs.\ \cite{FeFu2,FeFu3,GOli1}).

\blank{
We would like to consider the operator product expansion
\eq
\phi_{h_2} \cN_{p,q} \vec{p,q}
\;.
\en
}
{}From eqn.\ \ref{eq.l.c4} we have
\eq
\phi_{h_2}(z) L_m = \s(L_m) \phi_{h_2}(z)
\;,
\en
where
\eq
\s(L_m) = L_m - z^m( h_2(m+1) + z\partial)
\label{eq.si.def}
\;,
\en
is a representation of the Virasoro algebra of central charge $c$.
We also have a representation of the algebra of the $l_m$ operators,
\eq
\s(l_m) = l_m - z^m h_2
\;.
\en

The Verma module $M_{h_{p,q}}$ is equally well spanned
by the states
\eq
\tL_{i_1} \ldots \tL_{i_m} \vec{p,q}
\;,
\en
where $ i_j \leq i_{j+1} < 0$.
and so we can write the null vector at level $pq$ in $M_{h_{p,q}}$ as
$\cN_{p,q}\vec{p,q}$ where $\cN_{p,q}$ is a polynomial in $l_m$.
Then heuristically we have
\eq
\phi_{h_2}(z) \cN_{p,q} \vec{p,q}
=
\s(\cN_{p,q}) \phi_{h_2}(z) \vec{p,q}
\;.
\en
We can expand $\s(\cN_{p,q})$ in $z$ as
\eq
\s(\cN_{p,q})
=
\sum_{r=0}^{pq} S_{r} z^{-pq+r}
\;,
\en
where the dependence of $S_r$ on $h_2,p$ and $q$ is suppressed.
Note that $S_0 \equiv S_0(L_0)$ is a polynomial in $L_0$.

Correlation functions are defined initially as maps on Verma modules.
We would like null states to decouple so that we can consistently
define correlation functions on the irreducible representations of
the Virasoro algebra.
In particular we would like the three point function
\eq
\cev{h_1} \phi_2(z) \cN_{p,q}\vec{h_{p,q}}
=
S_0(h_1) \cev{h_1}  \phi_2(z)\vec{h_{p,q}}
=
0
\;.
\en
For non-vanishing three point functions we need $S_0(h_1) = 0$
where again the dependence of $S_0$ on $h_2,p$ and $q$ is suppressed.
The expression for $S_0(h_1)$
has been found for all $p,q,h_2$  by Feigin and Fuchs (see refs.\
\cite{FeFu1,FeFu2,FeFu3}) by an indirect method relying on
the factorisation of the Virasoro null vector expressions $\cN_{p,q}$
for special values of $t$, and directly for $p=1$ or $q=1$ by
Langlands (ref.\ \cite{Lang1}) and Bauer et al. (ref.\
\cite{BDIZ1,BDIZ2}),
and directly by Kent for all cases (ref.\ \cite{Kent4}).
{}From the ref.\ \cite{FeFu3} we then know all allowed non-vanishing
three point couplings.
However as we said at the start of this section, we are more used to
an operator product as some function
$\cF(z)$.
Let us assume from now on that we have a triple $\{h_1,h_2,h_{p,q}\}$
such that $S_0(h_1) = 0$.

Following Bauer et al.\ let us put
\eq
\cF(z) = \sum_{n\geq 0} f_n z^{y+n}
\quad,\qquad
\cJ(z) = \sum_{n\geq 0} j_n z^{y+n-pq} = \s(\cN_{p,q}) \cF(z)
\label{eq.fjdef}
\;,
\en
where $f_n,j_n \in M_{h_1}^{(n)}$.
Eqn.\ \ref{eq.fjdef} determines $j_n$ in terms of $f_n$ and $S_r$.

The following two results hold
\vskip 1mm \noindent I)\hskip 1cm%
If $S_{0}(h_1+n)\neq 0$ for $0<n < N$ it is possible to impose $j_n
= 0$ for $0\leq n <N$
by recursively defining
\eq
f_n = -\frac{1}{S_0} \sum_{r=1}^{pq} S_{r} f_{n-r}
\;.
\label{eq.fn.recur}
\en
Then the $f_n$ so defined obey the `descent equations'
\eq
 L_{m} f_n = (L_0 + m h_2 - h_{p,q})  f_{n-p}
\qquad
0\leq n<N, p>0
\;,
\label{eq.fn.desc}
\en
which state that the operator product
\eq
\phi_{h_2}(z) \vec{h_{p,q}} = \sum_{n\geq0} z^{y+n} f_n
\;,
\en
transforms covariantly under the Virasoro algebra.

\vskip 1mm \noindent II)\hskip 1cm%
If however $S_{0}(h_1+n) \neq 0, 0<n<N$, $S_{0}(h_1) = S_{0}(h_1+N) =
0$ then  we have
\eq
L_p j_N = 0
\qquad
p>0
\label{eq.gn.hw}
\;,
\en
i.e. $j_N$ is a (possibly identically
zero)  highest weight vector in  $M_{h_1}$ at level $N$  where
\eq
j_N = \sum_{r=1}^N S_{r} f_{n-r}
\;.
\en

\vskip 1mm

We prove these statements by induction on $n$ using the fact that $\s$ is
a representation of the Virasoro algebra, and the properties of
$\cN_{p,q}$.

\vskip 1mm \noindent (1) \hskip 1cm%
For $n=0$ we have $j_0 = 0, f_0 = \vec{h_1}$ and
so the results follow.

\vskip 1mm \noindent (2) \hskip 1cm%
Let us assume that
\eqq
&&S_0(h_1) = 0
	\qquad,\qquad\qquad
S_0(h_1+n) \neq 0 \;,\quad 0<n<M
	\nonumber\\
&&
L_m f_n = (L_0 + m h_2 - h_{p,q}) f_{n-m}  \; 0<n<M
\;.
\label{eq.conds}
\enn
The left ideal in $U(Vir)$, the universal enveloping algebra of the
Virasoro algebra, which annihilates the highest weight state
$\vec h_p$ is generated by $(l_m + z h_{p,q} \delta_{m,1}), m\geq 1$,
which is a restatement of the highest weight property
eqn.\ \ref{eq.hw.l}.
If we know that $\cN_{p,q}\vec{h_{p,q}}$ is a highest weight vector,
then
$(l_m + z h_{p,q} \delta_{m,1}) \cN_{p,q} \vec{h_{p,q}} = 0, m > 1$.
By considering the possible results of commuting $(l_m + z h_{p,q}
\delta_{m,1})$ through $\cN_{p,q}$ we find that
\eqq
(\tL_1 +z(h_{p,q}+pq)\cN_{p,q}
	&=& \cN'_{p,q} ( \tL_1 + z h_{p,q} )
\;,\nonumber\\
\tL_2 \cN_{p,q}
	&=& \cN''_{p,q} \tL_2
	+ \cN'''_{p,q} ( \tL_1 +z h_{p,q} )
\;, \nonumber\\
        &\vdots &\nonumber \\
\noalign{\vskip .2cm\noindent%
for some $\cN',\cN'',\cN'''$  polynomial in the modes $\tL_m$.
However, since $\s$ is a representation of the Virasoro algebra  with
the same central charge as $L_m$, we also have
}
(\s(\tL_1) + z(h_{p,q}+pq)\s(\cN_{p,q} )
	&=& \s(\cN'_{p,q} ) ( \s(\tL_1 ) + z h_{p,q} )
\;, \label{eq.s1}\\
\s(\tL_2 ) \s(\cN_{p,q} )
	&=& \s(\cN''_{p,q} ) \s(\tL_2 )
	+ \s(\cN'''_{p,q} ) ( \s(\tL_1 ) +z h_{p,q} )
\;, \label{eq.s2}\\
	&\vdots &\nonumber
\enn
If we put
\eq
\s(\cN'_{p,q}) = \sum_r S'_r z^{-pq+r}
\;,\quad
\s(\cN''_{p,q}) = \sum_r S''_r z^{-pq+r}
\;,\quad
\s(\cN'''_{p,q}) = \sum_r S'''_r z^{-pq+r}
\;.
\en
then we find that
\eq
S'_0 = S_0(L_0 + 1)
\quad,\qquad
S''_0 = S_0(L_0 + 2)
\;.
\en
Let us apply eqn.\ \ref{eq.s1} to $\cJ(z)$.
We have
\eqq
	&&
(\s(\tL_1) + z( h_{p,q}+pq)) \cJ(z)
\nonumber\\
	&=&
(\s(\tL_1) + z( h_{p,q}+pq)) \s(\cN_{p,q}) \cF(z)
\nonumber\\
	&=&
\s(\cN'_{p,q}) (\s(\tL_1) + z( h_{p,q})) \cF(z)
\nonumber\\
	&=&
S_0( L_0 + 1) ( L_1 f_M - (L_0 + h_2 - h_{p,q}) f_{M-1}) z^{y-pq+M}
+ O(z^{y-pq+M+1})
\;.
\enn
Comparing coefficients of $ z^{y-pq+M} $
we obtain
\eq
  L_1 j_M =S_0( h_1 + M)
	\left(
		L_1 f_M -  (L_0 + h_2 - h_{p,q}) f_{M-1}
	\right)
\;.
\en
If $S_0( h_1 + M) \neq 0$ then we have $j_M=0$ by the recursive
definition of $f_n$ and
\eq
L_1 f_M =  (L_0 + h_2 - h_{p,q}) f_{M-1}
\;.
\en
If  $S_0( h_1 + M) = 0$ then we have
\eq
L_1 j_M = 0
\;.
\en
We easily obtain the analogous results for $L_2$ by considering
$\s(\tL_2) \cJ(z) $

The null vector at level $M$ has the explicit expression
\eq
 j_M = \sum_{n1,\ldots n_j} S_{n_1} \frac{-1}{S_0} \ldots
	\frac{-1}{S_0} S_{n_j} \vec{h_1}
 \;,
\en
where the sum is over unordered partitions of $M$  into
numbers $1\leq n_i\leq pq$.
Thus we easily obtain eqns.\ \ref{eq.fn.desc} and \ref{eq.gn.hw} using
only properties of the representation $\s$. In
their paper Bauer et al.\ used a
neat inductive method based on the explicit expression for
$\s(\cN_{1,2})$ to prove these results.
\vskip 1mm

It is possible to apply these ideas to the fusion of
superconformal primary fields
\eq
M_2 \times M_3 \to M_1
\en
There are four possible assignments of the represntations
$i=1,2,3$ to NS and R
separately. The allowed fusions are

\begin{center}
\begin{tabular}{c c c|c}
    &   1   &   2   &   3   \\
\hline
(a) &   NS  &   NS  &   NS  \\
(b) &   NS  &   R   &   R   \\
(c) &   R   &   NS  &   R   \\
(d) &   R   &   R   &   NS
\end{tabular}
\end{center}
We shall only give details here for the
fusions of type (c), as this the method we shall use later to find
expressions for the Ramond null states

\subsection{Fusion type (c)}
\label{ssec6.c.0}

In this case we are not able to find a representation for
the Ramond algebra in differential operators as we could for the
Virasoro algebra in eqn.\ \ref{eq.si.def}.
 We could use the $2\times 2$ matrix  differential operator
representation suggested by eqn.\ \ref{eq.ns.c5}.
Instead
let us first  consider the combinations
\eqq
\tL_p &=& L_p - z L_{p-1}
\;,
\nonumber\\
\tG_p &=& G_p - z G_{p-1}
\;.
\label{eq.tltg.def}
\enn
These have the (anti)-commutation relations
\eqq
{}~[\tL_m,\tL_n ]
	&=&
(m-n) ( \tL_{m+n} - z \tL_{m+n-1} )
\cont+ \frac c6 m(m-1) \left( (m + mz^2 -2z^2)\delta_{m+n}
			- 2(m-1)\delta_{m+n-1}
		  \right)
\;,
\nonumber\\
{}~[\tL_m,\tG_n]
	&=&
  \left( \frac m2 -n \right) \tG_{m+n}
- z \left(\frac{m+1}2 -n\right) \tG_{m+n-1}
\;,
\\
\{ \tG_m,\tG_n \}
	&=&
  2\left( \tL_{m+n} - z \tL_{m+n-1} \right)
   \cont+ \frac{2c}3\left(
		   \left(
			\left(m^2-\frac 14 \right)
		   +z^2 \left( (m-1)^2 -\frac 14 \right)
		   \right) \delta_{m+n}
	       - 2 z \left( m -\frac 12 \right)^2 \delta_{m+n-1}
	        \right)
\nonumber
\;,
\enn
(note that these simplify rather for $z$=1).

Fusion gives the representation $\s$ of this algebra as
\eqq
\s( \tL_m) &= &\tL_m - z^m h_2
\nonumber\\
\s(\tG_m) &=& \epsilon_2 \tG_m
\;,
\label{eq.s.gr}
\enn
For simplicity let us suppose that $\epsilon=1$ in what follows.
The Verma module representation is still freely generated over this
algebra, and the ideal which annihilates the highest weight state
$\vec{\lambda}$
is generated by
\eq
( -\frac 1z\tL_p - (\lambda^2 - c/24)\delta_{p,1} ) \;,\, p\geq 1 ,
 \qquad\qquad
(  -\frac 1z\tG_p - \lambda \delta_{p.1}) \;,\, p\geq 1
\;.
\en
If $\lambda_3 = \lambda_{p,q}$ for $p,q \in \Bbb Z, p+q \in 2\Bbb Z+1$
then there is a highest weight vector $\cN\vec\lambda_3$ in $M_{\lambda_3}$
with $G_0$ eigenvalue $\tilde\lambda_3$ as given by eqn.\
\ref{eq.r.gonull}. The operator $\cN_{p,q}$
can be expressed as a polynomial in
the generators $\tL_m$ and $\tG_m$. Since $\{ \tL_m, \tG_m\}$ form a
closed algebra we have
\eqq
(  -\frac 1z\tL_1 - (\tilde\lambda_3^2 - c/24) ) \cN
&=&
\cN' ( -\frac 1z\tL_1 - (\lambda_3^2 - c/24) )
+
\cN'' ( -\frac 1z\tG_1 - \lambda_3)
\;,
\nonumber\\
(  -\frac 1z\tG_1 - \tilde\lambda_3^2  ) \cN
&=&
\tilde\cN' ( -\frac 1z\tL_1 - (\lambda_3^2 - c/24) )
+
\tilde\cN'' ( -\frac 1z\tG_1 - \lambda_3)
\;,
\label{eq.rr}
\enn
by considering the possible terms that can arise.
These relations automatically extend to the representation $\s$ since
it has the same central charge.

Let us now define as before
\eq
\cF(z) = \sum_n z^{y+n} f_n \;,\qquad
\cJ(z) = \sum_n z^{y+n - N} j_n \equiv \s(\cN) \cF(z)
\;,
\label{eq.cfcg}
\en
where $y = h_1 - h_2 - h_3$, $h_3 = \lambda_3^2 + c/24$,
$h_1 = \lambda_1^2 - c/24$, and finally $pq/2=\tilde\lambda_3^2 -
\lambda_3^2$ is the level in $M_3$ at which the null vector occurs.
We also define $S_r$ by
\eq
\s(\cN) = \sum_{r=0}^{pq/2} S_{r} z^{-{pq/2}+r}
\;,
\en
where $S_0$ is a polynomial in $G_0$, which we write $S_0(G_0)$.
Then we have
\eq
j_n = S_0 f_n + \sum_{r=1}^{pq/2} S_r f_{n-r}
\label{eq.z1}
\en
For the fusion to be possible  we require
$\cJ(z)$ to be null, and since $M_1^{(0)}$ is spanned by
$\vec{\lambda_1}$ we must have $j_0 \equiv 0$ and
\eq
S_0 f_0 = S_0( \lambda_1) \vec{\lambda_1} = 0
\;.
\en
If we have
\eq
S_0 (\lambda_1) = 0
\;,\quad
S_0 ( \pm( \lambda_1^2 + m)^{1/2}) \neq 0,\; 0<m<M
\;,
\en
then $S_0$ is invertible up to level $M-1$ in $M_1$ and we can
consistently set
\eq
j_m = 0
,\;
f_m = -\frac 1{S_0} \sum_{r=1}^N S_r f_{n-r}
,\;
0<m<M
\;.
\label{eq.ff}
\en

As in the Virasoro case we can prove the following 2 results.

\vskip 1mm \noindent I)\hskip 1cm%
Using eqns.\ \ref{eq.s.gr}
and \ref{eq.rr}  the states $f_n$ defined
recursively by eqn. \ref{eq.ff} obey the descent eqns.\
\eq
L_1 f_n = (L_0 - h_3 + h_2) f_{n-1}
\;,\quad
G_1 f_n = (G_0 - \lambda_3) f_{n-1}
\;,
\label{eq.gr.dec}
\en
which express the statement that $\cF(z)\sim
\phi_{h_2}\vec{\lambda_3}$ transforms covariantly under
the Ramond algebra.

\vskip 1mm \noindent II)\hskip 1cm%
If $S_0(\lambda_1)=0$,
$S_0(\pm(\lambda^2 + m)^{1/2}) \neq 0, 0<m<M$, and
$S_0(\bar\lambda)=0$ where $\bar\lambda^2 = \lambda_1^2 + M$ then the
state
\eq
\psi = \frac{ G_0 + \bar\lambda }{2\bar\lambda} j_M
\en
is a (possibly identically zero) highest weight state in
$M_{\lambda_1}^{(M)}$.

We now prove by I) and II) by induction on $m$.

\vskip .2cm
\noindent%
(1) For $m=0$ these equations are trivial.

\vskip .2cm
\noindent
(2)
Suppose the eqns.\ \ref{eq.gr.dec} are satisfied for
$ 0 \leq  m < M $.
We now consider $\s(\tL_1) \cF(z) $ and $\s(\tG_1) \cF(z)$.
The first of these yields
\eqq
\left( -\frac 1z\s(\tL_1) -h_3 \right) \cF(z)
&=&
( -\frac 1z L_1 + L_0 - h_3 + h_2) \sum_{n=0} f_n z^{y+n}
\nonumber\\
&=&
\sum_{n=0} \left(  (L_0 -h_3 + h_2) f_n - L_1 f_{n+1} \right) z^{y+n}
\;,
\enn
and the second
\eqq
\left( -\frac 1z\s(\tG_1) -\lambda_3 \right) \cF(z)
&=&
( -\frac 1z G_1 + G_0 - \lambda_3 ) \sum_{n=0} f_n z^{y+n}
\nonumber\\
&=&
\sum_{n=0} \left(  (G_0 -\lambda_3) f_n - G_1 f_{n+1} \right) z^{y+n}
\;.
\enn
We can write
\eq
S_0(G_0) = a(G_0^2) + b(G_0^2) G_0
\;,
\en
where $a,b$ are polynomials.
 We put
\eq
\begin{array}{rclrcl}
\s(N') &=& z^{-{pq/2}} ( S'_0 + O(z) )
&
\s(N'') &=& z^{-{pq/2}} ( S''_0 + O(z) )
\;,
\\
\s(\tilde N'') &=& z^{-{pq/2}} ( \tilde S'_0 + O(z) )
&
\s(\tilde N'') &=& z^{-{pq/2}} ( \tilde S''_0 + O(z) )
\;.
\end{array}
\label{eq.sp}
\en
Taking the leading term in $1/z$ in the $\s$ representations
of eqns.\ \ref{eq.rr} we get
\eq
L_1 S_0 = S_0' L_1 + S_0'' G_1
\;,\quad
G_1 S_0 = \tilde S_0' L_1 + \tilde S_0'' G_1
\;,
\en
and hence
\eq
\begin{array}{rclrcl}
S'_0 &=& a(G_0^2 + 1) + b(G_0^2 + 1) G_0
\;,
&
S''_0 &=& \frac 12 b(G_0^2 + 1)
\;,
\\
\tilde S'_0 &=& 2  b(G_0^2 + 1)
\;,
&
\tilde S''_0 &=& a(G_0^2 + 1) - b(G_0^2 + 1) G_0
\;.
\end{array}
\en
		%
Taking the coefficient of $z^{y + M - 1}$
in  $\s(\tL_1) \cJ(z) $ we get
\eqq
0 &=& (L_0 -h_{p,q} - {pq/2} + h_2) j_{M-1} - L_1 j_M
\nonumber\\
&=&
\sum_{r=0}^{pq/2} S'_r \left(  (L_0 -h_{p,q} + h_2) f_{M-r-1} - L_1 f_{M-r}
\right)
+
\sum_{r=0}^{pq/2} S''_r \left(  (G_0 -\lambda_{p,q} ) f_{M-r-1} - G_1 f_{M-r}
\right)
\nonumber\\
&=& S'_0  \left(  (L_0 -h_{p,q} + h_2) f_{M-1} - L_1 f_{M} \right)
+ S''_0 \left(  (G_0 -\lambda_{p,q} ) f_{M-1} - G_1 f_{M} \right)
\;.
	\\
\noalign{
\vskip 1mm
\noindent%
And similarly for $\s(\tG_1) \cG$ we get
\vskip 1mm
}
0 	&=&
	\tilde S'_0 \left(  (L_0 -h_{p,q} + h_2) f_{M-1} - L_1 f_{M} \right)
      + \tilde S''_0 \left(  (G_0 -\lambda_{p,q} ) f_{M-1} - G_1 f_{M} \right)
\;.
\enn
If $S_0$ is invertible for level $M$, it is easy to see
that the matrix
\eq
\pmatrix{ S'_0 & S''_0 \cr \tilde S'_0 & \tilde S''_0 }
\en
is invertible at level $(M-1)$ and so
\eq
\left(  (L_0 -h_{p,q} + h_2) f_{M-1} - L_1 f_{M} \right)
=
\left(  (G_0 -\lambda_{p,q} ) f_{M-1} - G_1 f_{M} \right)
=0
\en
This completes the inductive proof for all I) for level $M$.

When it does happen that $S_0$ is not invertible at level $M$, then
we can no longer choose $f_M$ such that $j_M$ is zero. However, in
this case we can find a (possibly identically zero) null state at
level $M$. Since $ G_0$ takes the two values
$\pm\sqrt{ \lambda_1^2 + M }$ on $M_1^{(M)}$,
let us call the value for which $S_0(G_0)$ is zero $\bar\lambda$.
We can project onto the $G_0$ eigenspaces $\pm\bar\lambda$ with
\eq
\pi_{\pm} = \frac{ \bar\lambda \pm G_0 }{2 \bar \lambda}
\;,
\en
then we have
\eq
\pi_+ S_0 \Bigg|_{M_1^{(M)}} = 0
\;.
\en
Eqn.\ \ref{eq.z1} becomes
\eq
\pi_+ j_M =
\sum_{r=1}^{pq/2} S_r f_{M-r}
\;.
\en
We now act with
$ \s(-\frac 1z\tL_1)  -h_{p,q} $, and get
\eqq
L_1 (G_0 + \bar\lambda) &=&
(G_0 + \bar\lambda) S_0(\bar\lambda)
	(L_1 f_M - (L_0 - h_{p,q} + h_2) f_{M-1})
\cont+
\frac 12 S_0(\bar\lambda)
	(G_1 f_M - (G_0 - \lambda_{p,q} ) f_{M-1})
\nonumber\\
&=& 0
\enn
If we denote the state $(G_0 + \bar\lambda) j_M$ by $ \psi$, then we have
shown
\eq
G_0 \psi = \bar\lambda \psi
\;,\quad
L_1 \psi = 0
\;.
\en
and hence $\psi$ is a Ramond highest weight state. The highest weight
state $\psi$ has the explicit form
\eq
\psi = \frac{ G_0 + \bar\lambda}{2\bar\lambda}
\sum_{\{n_1,\ldots n_m} S_{n,1} \frac{-1}{S_0} S_{n_2} \ldots
\frac{-1}{S_0} S_{n_m} \vec{\lambda_1}
\;,
\en
where the summation is over all partitions of $M$ into numbers
$1 \leq n_i \leq pq$

\subsection{Comments}
\label{ssec6.com}

In the both the Virasoro and Superconformal cases it may happen
the fusion procedure breaks down at some level $N$ where there are no
highest weight vectors in $M_{h_1}$. This will occur when
representations of which the conformal weights differ by an integer
occur in the set of allowed $h_1$.
In this case however, we have proven that the states $j_N$ and $\pi_+ j_N$
respectively must vanish identically.

A  related phenomenon
occurs when one considers
differential equations for four
point functions. A four point function of primary fields of the
Virasoro algebra
\eq
\cev{ \phi_a} \phi_b(1) \phi_c(z) \vec{\phi_d}
\en
will satisfy a differential equation in $z$ if there is a null vector in
$M_{d}$. This equation has regular singular points at $0,1,\infty$.
Expanding in $z$ about $0$ we can find the indicial equation
describing the behaviour of power series solutions to the differential
equation. If any two the roots of the indicial equation differ by an
integer then this will generally  signal the presence of logarithmic
solutions to
the differential equation.
However, in the cases where this happens the equation is such that
there are no logarithmic solutions, essentially following from our
argument above that $j_N$ vanishes.
An alternative proof that there are no logarithmic solutions could come
from constructing the actual chiral vertex  operators $\phi(z)$ and
showing that they generate a sufficient number of solutions to the
differential equations. This has been shown for the Virasoro minimal
model cases ($c=1 - 6/(m(m+1)), h=h_{p,q}$) by T.\ Loke (\cite{Loke1})
exploiting the
coset construction (see refs.\ \cite{GKOl1,GKOL2}).

\vf

\section{Operator product expansions  for simple null fields}
\label{sec.ope.null}

We shall consider the simplest cases of a null vector in the Verma
module 
and see whether we are
able to find any cases where the fusion procedure breaks down and we
can construct null states in $M_1$.

We shall confine ourselves to considering the cases
where the null vector in $M_3$ takes the simplest form in the two
cases of Neveu-Schwarz and Ramond.
These are:
\vskip .2cm
\noindent In the NS sector, the representation (1,3) has a null state
$\cN_{1,3}\vec{1,3}$, where
\eq
\cN_{1,3} = G_{-3/2} - \frac 1t L_{-1}G_{-1/2}
\quad,\qquad
h_{1,3} = t - 1/2
\;.
\en 

\noindent In the R sector, the representation (1,2) has a null state
$\cN_{1,2}\vec{1,2}$, where
\eq
\cN_{1,2} = L_{-1} +\sqrt{\frac t2} G_{-1}
\quad,\qquad
h_{1,2} = \frac{3t}8 - \frac 3{16}
\quad,\qquad
\lambda_{1,2} = \frac{ 1 - 2t}{2\sqrt{2t}}
\;.
\en

\subsection{Case (a)}
\label{ssec6.a}

This has been studied before (see e.g.\ refs.\
\cite{BKTe1,Eich1,SSta1})
and results in the well known fusion rule
\eq
(p,q) \times (1,3) \to
	\cases{ (p,q+2), (p,q-2) & even \cr
		(p,q)		 & odd }
\en
This was the method used by Benoit and Saint-Aubin (ref.\
\cite{BStA3}) and Hwang et al.\ (ref.\ \cite{HZZh1}) to produce
formulae for the general Ramond vectors, using a superfield formalism.
They also obtained explicit expressions for the analogue in this
formalism of the numbers $S_{n,0}$ of subsecn.\ \ref{ssec6.vir}.

\subsection{Case (b)}
\label{ssec6.b}

We consider the OPE
\eq
\phi_{\lambda_2}(z) \vec{\lambda_{1,2}} \in M_{h_1}
\;.
\label{eq.b.fu}
\en
We can write $\cN_{1,2}$ as
\eq
\cN_{1,2} =
\tL_{-1} + \frac{h_{1,2}}{z}
 + \sqrt{\frac t2}\left( K_{-1} + \frac{\lambda_{1,2}}{2} \right)
\;.
\en
Using
\eq
\s(K_m(z)) =
   \epsilon \left( \lambda z^{m-1}  - z^{-1/2} K^+_{m+1/2}(z) \right)
\;,
\label{eq.sk}
\en
we have that
\eqq
\s(\cN_{1,2}) &=&
\frac 1z
\left(
h_{1,2} - L_0  + \epsilon\sqrt{\frac t2}  \frac{\lambda_2}{z}
\right)
+
L_{-1}
-
\epsilon\sqrt{\frac t2}  K^+_{-1/2}(z)
\nonumber\\
&=& \frac {S_0}z + O(z^{-1/2})
\;.
\label{eq.b.s0}
\enn
and so $S_0 f_0 = 0$ implies $ S_0 f_n \neq 0 $ for $n \neq 0$
and hence the operator product expansion states $f_n$ can be  defined
recursively for all $n$ and this method does not produce any new null
states in $M_1$ by the means of section \ref{sec.ope}.

\subsection{Case (c)}
\label{ssec6.c}

We consider the case $\phi_{h_2}$ a bosonic Neveu-Schwarz field for
simplicity. 
We have
\eq
\cN_{1,2} =
\tL_{-1} + \frac{h_{1,2}}z
+
\sqrt{\frac t2}\left( \tG_{-1} + \frac{\lambda_{1,2}}z \right)
\;.
\en
If we put $h_2= h_{p,0}$ with $p$ not necessarily integral, we get
\eqq
\s(\cN_{1,2}) &=&
-\frac 1z(G_0 -\lambda_{p,1})(G_0 - \lambda_{-p,1})
+
\left( L_{-1} + \sqrt{\frac t2}G_{-1} \right)
\nonumber\\
&=& \frac{S_0}z + S_1
\;.
\label{eq.f1}
\label{eq.c.s0}
\enn
{}From eqn.\ \ref{eq.f1} and eqns.\ \ref{eq.cfcg} we obtain
\eq
g_n =
\left( L_{-1} + \sqrt{\frac t2}G_{-1} \right) f_{n-1}
-
\frac 1z(G_0 -\lambda_{p,1})(G_0 - \lambda_{-p,1}) f_n
\;.
\label{eq.f2}
\en
We firstly require that $f_0 = \vec{\lambda_1}$ and that $g_0 = 0$ as
there can be no null state at
level 0 (the Verma module at level zero is one dimensional spanned by
the highest weight state). Thus the allowed fusions are
\eq
(p,0) \times (1,2) \to (p,1), (p,-1)
\;.
\label{eq.fu1}
\en
We can then invert
\eq
\Delta \equiv (G_0 -\lambda_{p,1})(G_0 - \lambda_{-p,1})
\;,
\label{eq.delta}
\en
at level $n$ for for
\eq
G_0^2 = n + \lambda_{\pm p,1}^2 \neq \lambda_{\mp p,1}
\;,
\label{eq.cond.c}
\en
and obtain
\eq
j_n \equiv 0 \quad,\qquad
f_n = \frac 1\Delta (L_{-1} + \sqrt{\frac t2}) f_{n-1}
\;.
\en
Condition (\ref{eq.cond.c}) is not met for
\eq
n = \pm p/2
\;,
\en
and hence we can solve for $j_n\equiv 0$ unless
\eq
\lambda_1 = \lambda_{2n,1}
\quad,\qquad
h_2 = h_{2n,0}
\;,
\label{eq.cond.fail}
\en
where $n$ is a positive integer. In the case
(\ref{eq.cond.fail}) we find
that the fusion fails at level $n$ and the (possibly zero) null state
we produce has $G_0$ eigenvalue $\lambda_{-2n,1}$ as predicted by Cohn
and Friedan in ref\ \cite{CFri1}.

\subsection{Case (d)}
\label{ssec6.d}

We consider the OPE
\eq
\phi_{\lambda_{2}}(z) \vec{ h_{1,3}} \in M_{ \lambda_1 }
\;.
\en
We have
\eq
\cN_{1,3}
=
K_{-3/2} + \frac 1{2z}K_{-1/2}
- \frac1t
	\left( \tL_{-1} + \frac{h_{1,3}+1}z \right)
	K_{-1/2}
\Bigg|_{G_p=0, p>0}
\;,
\en
and hence if we put $\lambda_2 = \lambda_{p,0}$ with $p$ not
necessarily an integer we get
\eqq
\s(\cN_{1,3})
&=&
-\frac{\epsilon_2}{t z^{3/2}}
		(  G_0 - \lambda_{p,-2} )
		(  G_0 - \lambda_{-p,0} )
		(  G_0 - \lambda_{p,2}  )
+ O(z^{-1/2})
\nonumber\\
&=& S_0(G_0) z^{-3/2} + O(z^{-1/2})
\;.
\label{eq.d.s0}
\enn
If we define
$\cF(z), \cG(z) = \s(\cN_{1,3}) \cF(z)$, then
we have a possible failure of the recurrence relation for $f_n$ if
$S_0$ has a zero eigenvalue.
Let us consider the three allowed cases $S_0(\lambda_1)=0$ in turn.

\subsubsection{ $\lambda_1 = \lambda_{-p,0}$}

The condition for failure of the recurrence at level $n$ is
\eq
G_0^2 -\lambda_{p,\pm 2}^2
=
\lambda_{p,0}^2 - \lambda_{p,\pm 2} + n
=
n - t/2 \mp p/2
= 0
\;.
\en
Hence we have a failure for $p=\pm(2n -t)$, $\lambda_1 =
\lambda_{\mp 2n,\mp 1}$  in which case failure is at level $n$ with
$G_0$
eigenvalue $\lambda_{\pm 2n,\mp 1}$.
This gives the fusion
\eq
(\pm 2n, \pm 1) \times (1,3) \to (\mp 2n, \mp 1)
\;.
\label{eq.fu2}
\en

\subsubsection{ $\lambda_1 = \lambda_{p,\pm 2}$}

The condition for failure of the recurrence relation for $f_n$
is
\eq
\lambda_{p,\pm 2}^2 -  \lambda_{-p,0}^2 + n =0
\;,
\label{eq.van.1}
\en
or
\eq
\lambda_{p,\pm 2}^2 - \lambda_{p,\mp2}^2 + n =0
\;.
\label{eq.van.2}
\en
Taking eqn.\ \ref{eq.van.1} first we then have
$p = \pm(2n + t)$ for $n$ a positive integer, and the fusion is
\eq
(\pm2n,\mp1) \times (1,3) \to (\pm 2n,\pm 1).
\en
The fusion fails at level $n$ and the $G_0$ eigenvalue is as predicted
in \cite{CFri1}.
If we have eqn.\ \ref{eq.van.2} then
we have $p= \pm n$ (we need $n$ an even integer here)
and the fusion is
\eq
(\pm n,0) \times (1,3) \to (\pm n,\pm 2)
\;,
\label{eq.fu3}
\en
where the fusion fails at level $n$ and the $G_0$ eigenvalue is as predicted
in \cite{CFri1}.

\subsection{Comments}
\label{ssec6.com2}

Benoit and St.\ Aubin suggested the fusion of subsecn.\  \ref{ssec6.b}
as a possible  way of producing Neveu-Schwarz null vectors.
Unfortunately this does not prove possible in the direct manner of
section  section \ref{sec.ope}. However, one could certainly use  the
allowed fusions
\eq
(p,q) \times (1,2) \to (p,q+1)
\quad,\qquad
(p,q) \times (2,1) \to (p+1,q)
\;,
\en
to generate all null vectors if one is prepared to use knowledge of
the null vectors in both representations in the manner suggested by
Bauer et al.\ in ref.\ \cite{BDIZ1}.

We also see in the fusion
(\ref{eq.b.fu}) that the choice of the `continuation
parameter' $\epsilon$ in eqn.\ \ref{eq.r.c4} affects the fusion
algebra of the theory.
As we stated in subsection \ref{ssec3.r}, the choice of  $\epsilon$
is equivalent to a choice of
adjoint. If we denote the vacuum representation (1,1) by $\id$
then
\eqq
(1,2) \times (1,2) &\to& \cases{
	(1,3) & $\eta=1$ \cr
	\id   & $\eta=-1$}
\nonumber\\
(-1,-2) \times (1,2) &\to& \cases{
	\id   & $\eta = 1$ \cr
	(1,3) & $\eta=-1$ }
\enn
We are not sure how to resolve this ambiguity.
It seems clear that both choices will lead to the definition of an
operator product which satisfies the descent equations. However
if one then attempts to define an associative operator product algebra
some choice may be imposed.
Some possibilities are:
the choice of $\epsilon_\lambda$ is free and may vary freely with the sector
in which $\phi_\lambda$ acts;
the values of $\epsilon_\lambda$ for different sectors may be forced having
chosen a value for $\phi_\lambda$ acting in the vacuum sector, or
giving a choice of adjoint;
the only consistent procedure is to include $(-1)^F$ and consider
representations of the extended algebra which have (generically) a
two-dimensional highest weight space;
one must go to a full theory with both holomorphic and
anti-holomorphic coordinates to obtain an associative operator product
algebra.

Mussardo et al.\ used the $(2,1)$ null vector to obtain the fusion
rules for this field from the requirement that it decoupled from three
point functions. They used an extended algebra comprising the Ramond
or Neveu-Schwarz algebra and the operator $(-1)^F$, and consequently
they had a different highest weight structure in the Ramond sector,
the highest weight spaces of this extended alegbra typically being
two-dimensional. \vf

\section{Explicit formulae for the $(2n,1)$ null vectors}
\label{sec.p1}

We wish to construct the $(2n,1)$ null states. There are two methods
open to us by considering fusions of the simplest Ramond and
Neveu-Schwarz null states. These are from (\ref{eq.fu1})
\eq
(2n,0) \times (1,2) \to (2n,1)
\;,
\label{eq.fuse.c}
\en
and from (\ref{eq.fu2},\ref{eq.fu3})
\eq
(-2n,-1) \times(1,3) \to (2n,1)
\quad,\qquad
(2n,-1) \times (1,3) \to (2n,1)
\;.
\label{eq.fuse.d}
\en
We have the expressions for $S_0$ in these  cases given by eqns.\
\ref{eq.c.s0} and \ref{eq.d.s0}. These are quadratic and
cubic in $G_0$ for fusions
(\ref{eq.fuse.c}) and (\ref{eq.fuse.d}) respectively.
Since we must invert $S_0$
to define the sequence $f_n$ recursively  we
choose the simpler case (\ref{eq.fuse.c}) to generate null vectors in
the R sector.

\subsection{$(2n,0) \times (1,2) \to (2n,1)$}
\label{ssec7.r}

{}From eqn.\ \ref{eq.f2}
we define
\eq
f_m =
\frac 1\Delta \left( L_{-1} + \sqrt{\frac t2}G_{-1} \right) f_{m-1}
\;
0<m<n
\quad,
\qquad
f_0 = \vec{p,1}
\;,
\en
where $\Delta$ is as in eqn.\ \ref{eq.delta}.
At level $n$ eqn.\ \ref{eq.f2}
reads
\eq
-j_n =
\left( L_{-1} + \sqrt{\frac t2}G_{-1} \right) f_{n-1}
-
 (G_0 -\lambda_{2n,1})(G_0 - \lambda_{-2n,1}) f_n
\;.
\en
Since at level $n$ in the space $M_{\lambda_{2n,1}}$ we have $G_0^2 =
\lambda_{-2n,1}^2$, if we project onto the
$\lambda_{-2n,1}$  eigenspace of $G_0$ we obtain
\eq
-(G_0 + \lambda_{-2n,1}) j_n
=
(G_0 + \lambda_{-2n,1}) \left( L_{-1} + \sqrt{\frac t2}G_{-1} \right) f_{n-1}
\;.
\label{eq.psi}
\en
Since the contribution from $f_n$ decouples here we cannot arrange
$j_n \equiv 0$ by  taking $f_n$ suitably.  In subsection
\ref{ssec6.c.0} we analysed this case carefully and proved  that
\eq
(G_0 + \lambda_{-2n,1}) \left( L_{-1} + \sqrt{\frac t2}G_{-1} \right) f_{n-1}
\en
is null, and so a highest weight state in $M_{(2n,1)}^{(n)}$.

We can in fact re-write this state in a very nice form.
Firstly we note that
\eq
\frac 1\Delta
= 
	\frac 1{(G_0 -\lambda_{2n,1})(G_0 - \lambda_{-2n,1})}
= 
\frac{(G_0 +\lambda_{2n,1})(G_0 + \lambda_{-2n,1})}
		{(G_0^2 -\lambda_{2n,1}^2)(G_0^2 - \lambda_{-2n,1}^2)}
\;,
\en
and that, at level $m$,  $G_0^2$ takes the value
$\lambda_{2n,1}^2 + n$. So, putting
\eq
D =
 \left( L_{-1} +  \sqrt{\frac t2}G_{-1} \right)
\quad,\qquad
X = (G_0 + \lambda_{-2n,1}) D  (G_0 +\lambda_{2n,1})
\;,
\en
we can re-write eqn.\ \ref{eq.psi} as
\eqq
	(G_0 + \lambda_{-2n,1}) D f_{n-1}
&=&	\frac 1{(n-1)(-1)}
	(G_0 + \lambda_{-2n,1}) D
	(G_0 +\lambda_{2n,1})(G_0 + \lambda_{-2n,1}) D
	f_{n-2}
\nonumber\\
&=&	\frac 1{2(n-1)(n-2)}
	X X (G_0 + \lambda_{-2n,1}) D
	f_{n-3}
\nonumber\\
&\vdots&
\nonumber\\
&=&  \frac 1{ ( (n-1)!)^2} \;
X X
\ldots
X
(G_0 + \lambda_{-2n,1}) D f_0
\;.
\enn
Unlike the Virasoro $(p,1)$ and NS $(2p+1,1)$ expressions, we can
clear the denominator
in this expression, and so rescaling we obtain the null state
$ \cN_{2n,1} \vec{2n,1}$
at level $n$ in $M_{\lambda_{2n,1}}$ as
\eq
\cN_{2n,1} = X^n
\;.
\label{eq.p1.null}
\en
It is noteworthy that this is
a completely factorised expression, but includes the modes $G_0$.
The proof that this is indeed a highest weight vector follows from the
theory outlined in section \ref{sec.ope}.

\subsection{A second formula for $\cN_{2n,1}$}
It
is possible to eliminate these modes at the expense of replacing the
formula for $\psi$ by a $2\times 2$ matrix $\Psi$ all the entries of which
are proportional and are highest weight vectors at level $n$ in
$M_{\lambda_{2n,1}}$.
\eq
\Psi =
N_{n} D N_{n- 1} \ldots  D N_0
|2n,1\rangle
\;,
\en
where
$D$ and $N_m$ are the $2\times 2$ matrices
\eqq
D &=&
\pmatrix{
L_{-1} + \sqrt{\frac t2} G_{-1} & 0 \cr 0 &  L_{-1} - \sqrt{\frac t2} G_{-1} }
\;,
\nonumber\\ 
N_m
&=&
\pmatrix{ m - (2n+t+1)/4 & -t/2 \cr
m - ((2n+t)^2 - 1)/(8t) & m - (2n+t-1)/4 }
\;.
\enn
This can be deduced directly by a fusion argument using the
$2\times 2$ matrix representation of the Ramond algebra suggested by
eqn.\ \ref{eq.ns.c5}.
This representation of the null vector has some nice properties in
that: firstly it can be seen that the expression is not identically
zero; secondly one can express the $G_0$ action as follows.
Let us define a series of vectors, which are in fact the operator
product expansion states obtained by using the $2\times 2$ matrix
representation, as follows:
\eq
F_m
=
N_{m} D F_{m-1}
\;,\;\; 0 < m \le n\,,\;\; F_0 = N_0  \vec{2n,1}
\en
Then we have, for $0 \leq m \leq n$,
\eqq
G_0 F_m
&=&
{1\over{ 2 \sqrt{2 t}}}
\pmatrix{ 1-2t & 2t \cr
	2 - 2n + {{4n^2 - 1}\over{2t}} - {{3t}\over 2} + 4m & 2t - 1}
F_m
\;,
\nonumber\\
L_1 F_m
&=&
m(m-n)
\pmatrix{ {{4n^2 - 1}\over{4t}} - {{2n+1}\over 4} - {t\over2} +m & 0 \cr
0 &  {{4n^2 - 1}\over{4t}} - {{2n-1}\over 4} - {t\over2} +m } F_{m-1}
\;.
\enn
These results can be easily obtained by induction, and show that
indeed $\Psi$ is a matrix of null vectors.

To obtain the expressions for the null vectors
in representations $(-2n,-1),(1,2n)$ and $(-1,-2n)$
it is only necessary to change $\sqrt t \to -\sqrt t$,
$ \sqrt t \to -1/\sqrt t$ and $ \sqrt t \to 1/\sqrt t$ respectively in
the expression we have given.\vf

\section{Formulae for the $(2n-1,2p)$ Ramond null vectors}
\label{sec.pq}

All we need now to produce general $(2n-1,2p)$ null vectors is to find
$S_0(G_0)$ for the fusions
\eq
h_{0,2p} \times \lambda_{2n,1} \to \lambda'
\;,
\en
for $p$ not necessarily integral.
Using the explicit expressions of $\cN_{2n,1}$ obtained in the
previous section, we can examine $S_0$ for small values of $n$, and
find
\eq
S_0(\lambda')
	 =
 2^{3 - 4n} t \prod_{j=1}^n
\left( ( 2\sqrt{2t}\lambda' - (-1)^{n+j}(2j-1))^2 - p^2 \right)
\;.
\label{eq.so.fact}
\en
This means that the allowed fusions are
\eq
(0,2p) \times (2n,1) \to
\left\{
\matrix{
( 2n-1,2p)	&,&\;\; ( 2n-1,-2p) \cr
(-2n + 3,2p)	&,&\;\; (-2n + 3,-2p) \cr
( 2n -5,2p)	&,&\;\; ( 2n-5,-2p) \cr
\vdots		& &\vdots \cr
((-1)^{n+1},2p)
		&,&\;\; ( (-1)^{n+1},-2p)
}
\right.
\en
Of these we could use any of the pairs of fusions for $2p$ integral to
generate null vectors, in particular the first pair.
Thus we can generate the null vectors in the R repreentation
\eq
(2n-1,2p)
\;,
\en
from a fusion
\eq
(0,2p) \times (2n,1)
\;,
\en
exactly as we laid out in section \ref{sec.ope}.
This can easily be written out in an explcit form, but as it is rather
cumbersome we shall not include it here.

However, we need to say again that at present eqn.\ \ref{eq.so.fact}
is only based on explicit calculations of $S_0$  small values of $n$.
We expect that this factorisation of $S_0$  can be easily proven directly
using some variant of the
matrix method of Bauer et al., and it certainly can be proven
indirectly along the lines of the Virasoro factorisation by Feigin and
Fuchs or by Kent.
\vf

\section{Conclusions}
\label{sec.conc}

In this paper we have examined the properties of fusion of
representations of the superconformal algebra in both Ramond and
Neveu-Schwarz sector. In section \ref{sec.ope} we showed that
using only representation theoretic arguments the presence of null
vectors in a Verma module  enables one to restrict the allowed
non-vanishing three point functions, find a sequence of states  which
satisfy the descent equations, and in certain cases find expressions
for the null vectors in new representations spaces.

We applied this theory to the construction of the null vectors in the
$(2n,1)$ Ramond representations which turned out to have a
particularly simple form. We were then able to conjecture fusions
which would produce the general Ramond null vector.

In the process we found that the fusion of Ramond fields has some
features which are not  very well detailed in the literature. In
particular the fusion rules of representations of the unextended
Ramond algebra, without $(-1)^F$ seem to have new features which
deserve investigation.

We also found that it was necessary to postulate the analytic continuation of
operator products of $G(z)$ and Ramond fields, and that there were (at
least) two natural definitions which seemed to be linked to the choice
of inner product on the Hilbert space. However, these choices seem
unavoidable in a chiral theory, and it is to be hoped that the
imposition of locality in a full  conformal field theory (with fields
having both holomorphic and anti-holomorphic coordinates)  would
remove the ambiguities.

\section*{Acknowledgements}

I am very grateful to
P.~Bowcock,
M.\ Chu,
M.~D\"orrzapf,
P.\ Goddard,
A.~Kent and T.\ Loke for
discussions on many aspects of fusion and superconformal field theory.
This work was supported by a research fellowship from  St.~John's
College, Cambridge.

\vf


\end{document}